\documentstyle[aps]{revtex}

\newcommand{\ket}[1]{|{#1}\rangle}
\newcommand{\bra}[1]{\langle{#1}|}
\newcommand{\kets}[2]{|{#1}\rangle_{{}_{\!\!{#2}}}}
\newcommand{\bras}[2]{{}_{{}_{{#2}\!\!}}\langle{#1}|}
\newcommand{\slb}[2]{{#1}^{({#2})}}

\newcommand{\cH}{{\cal H}}
\newcommand{\cG}{{\cal G}}

\newcommand{\defeq}{\doteq}

\newcommand{\trace}{\mbox{tr}}

\newcommand{\mod}{\mbox{mod}}
\newcommand{\lowertr}{\mbox{lwtr}}
\newcommand{\diag}{\mbox{diag}}

\newtheorem{Theorem}{Theorem}
\newtheorem{Corollary}[Theorem]{Corollary}
\newtheorem{Lemma}[Theorem]{Lemma}
\newtheorem{Problem}[Theorem]{Problem}

\newcommand{\proof}{\paragraph*{Proof.}}

\newcommand{\qed}{\hspace*{\fill}\rule{2.5mm}{2.5mm}%
\vspace*{8pt}\par}

\newcommand{\phead}[1]{\par\noindent{\bf #1}}

\begin{document}
%\twocolumn

\title{Quantum Computation and Quadratically Signed Weight Enumerators}

\author{E. Knill$\;^*$, R. Laflamme
\thanks{E-mail addresses: {\tt knill@lanl.gov, laflamme@lanl.gov}}}
\address{
Los Alamos National Laboratory, MS B265, Los Alamos, New Mexico 87545}

\date{September 1999}
\maketitle

\begin{abstract}
We prove that quantum computation is polynomially equivalent to
classical probabilistic computation with an oracle for estimating the
value of simple sums, {\em quadratically signed weight
enumerators\/}. The problem of estimating these sums can be cast in
terms of promise problems and has two interesting variants. An oracle
for the unconstrained variant may be more powerful than quantum
computation, while an oracle for a more constrained variant is
efficiently solvable in the one-bit model of quantum
computation. Thus, problems involving estimation of quadratically
signed weight enumerators yield problems in BQP (bounded error
quantum polynomial time) that are distinct from
the ones studied so far, include a canonical BQP complete problem,
and can be used to define and study complexity classes and their
relationships to quantum computation.
\end{abstract}

\section{Introduction}

It is widely believed that quantum computers are more efficient than
classical deterministic or probabilistic computers.  For example,
there is an efficient algorithm for factoring integers on a quantum
computer, while no such algorithm is known for classical
computers~\cite{shor:qc1994a,shor:qc1995a}.  Unlike numerous other
models more efficient than traditional computation, quantum
computation appears to be robustly implementable using reasonable
physical
devices~\cite{shor:qc1996a,aharonov:qc1996a,kitaev:qc1997a,knill:qc1997a,knill:qc1998a}.

To better understand the power of quantum computers, it is desirable
to find specific problems to which the problem of simulating a quantum
computer on a classical computer can be reduced.  In principle, such
problems can be extracted from the representation of the amplitudes of
the desired answer of a quantum algorithm as a sum over paths of
transition amplitudes. (The sum is over all possible evolutions of the
computational states consistent with the steps of the quantum
algorithm.) This representation can be used to prove that quantum
computers can be simulated on classical computers with exponential
overhead in time and polynomial overhead in
space~\cite{bernstein:qc1997a,aharonov:qc1998a,cleve:qc1999b}.  The
resulting problems can be simplified by using the fact that transition
amplitudes can be restricted to a small set of rational
numbers~\cite{bernstein:qc1997a}. However, these path sums are still
too general for use as canonical problems whose solutions suffice for
efficient simulation of quantum computers.  Furthermore, it is not
clear how to modify the path sums to represent related computational
models such as the one-bit model of quantum
computation~\cite{knill:qc1998c}.  This model differs from standard
quantum computation in that the initial state is random except for one
quantum bit, and measurement is destructive. The goal of this paper
is to remedy this situation by relating both the standard and the
one-bit model of quantum computation to problems of estimating certain
sums related to weight generating functions for binary codes.  Some of
these estimation problems can be cast as promise problems with the
property that oracles for these problems can be used to efficiently
predict the answers of quantum algorithms. Conversely, since there are
efficient quantum algorithms and one-bit quantum algorithms for
solving such promise problems, they define a new class of problems in
BQP that are apparently hard for classical computation.

\phead{Quadratically Signed Weight Enumerators.}
A general quadratically signed weight enumerator is of the form
\begin{eqnarray}
S(A,B,x,y) &=& \sum_{b:Ab=0}(-1)^{b^TBb}x^{|b|}y^{n-|b|},
\end{eqnarray} 
where $A$ and $B$ are be $0$-$1$-matrices with $B$ of dimension $n$ by
$n$ and $A$ of dimension $m$ by $n$.  The variable $b$ in the summand
ranges over $0$-$1$-column vectors of dimension $n$, $b^T$ denotes the
transpose of $b$, $|b|$ is the {\em weight\/} of $b$ (the number of
ones in the vector $b$), and all calculations involving $A$, $B$ and
$b$ are modulo $2$. The absolute value of $S(A,B,x,y)$ is bounded by
$(|x|+|y|)^n$. In general, one can consider the computational problem
of evaluating these sums.  Here we consider the following cases, which
will be related to quantum computation:

\begin{Problem}
Given that $k$ and $l$ are positive integers, evaluate
$S(A,B,k,l)$.
\label{problem:exact_eval}
\end{Problem}

Problem~\ref{problem:exact_eval} is in the class
$\#P$~\cite{papadimitriou:qc1994a}.

\begin{Problem}
Given that $k$ and $l$ are positive integers and the promise
$|S(A,B,k,l)| \geq (k^2+l^2)^{n/2}/2$,
determine the sign of $S(A,B,k,l)$.
\label{problem:estimate_gen}
\end{Problem}

The next two problems require that $A$ is square.
Let $\lowertr(A)$ denote lower triangular part of $A$, which is the
matrix obtained from $A$ by setting to zero all the entries on or
above the diagonal.  Let $\diag(A)$ denote the diagonal matrix whose
diagonal is the same as that of $A$. $I$ denotes the identity
matrix. For matrices $C$ and $D$ with the same number of columns,
$[C;D]$ denotes the matrix obtained by placing $C$ above $D$.

\begin{Problem}
\label{problem:estimate_dqc}
Given that $\diag(A)=I$,
$k$ and $l$ are positive integers,
and the promise $|S(A,\lowertr(A),k,l)| \geq (k^2+l^2)^{n/2}/2$,
determine the sign of $S(A,\lowertr(A),k,l)$.
\end{Problem}

\begin{Problem}
\label{problem:estimate_dqc1}
Given that $\diag(A)=I$, $k$ and $l$ are
positive integers and the promise $|S([A;A^T],\lowertr(A),k,l)| \geq
(k^2+l^2)^{n/2}/2$, determine the sign of $S([A;A^T],\lowertr(A),k,l)$.
\end{Problem}

We will show that Problem~\ref{problem:estimate_dqc} is BQP complete,
so that classical probabilistic computation with an oracle for this
problem is polynomially equivalent to quantum
computation. Problem~\ref{problem:estimate_dqc1} is solvable
efficiently using a one-bit quantum algorithm.  In the last two
problems, the integers $k$ and $l$ can be restricted to $4$ and $3$,
respectively, without affecting their hardness with respect to
polynomial reductions (using classical deterministic algorithms).

\section{Models of Quantum Computation}

An easy-to-use model of quantum computation consists of a classical
random access machine (RAM)~\cite{papadimitriou:qc1994a} with access
to any number of addressable {\em quantum bits} (qubits) that are
initially in the state $\ket{0}$. The qubits can be manipulated by one
of a finite set of {\em quantum gates\/} and by {\em measurement\/}.
This model is called the quantum random access machine (QRAM).
For introductions to the basic notions of quantum computing,
see~\cite{aharonov:qc1998a,cleve:qc1999b}. 

The basic states of qubit
$A$ are denoted by $\kets{0}{A}$ and $\kets{1}{A}$. These
are elementary {\em ket\/} symbols. The basic states
of a collection of qubits are obtained by formally multiplying the
basic states of each qubit. For example,
$\kets{0}{A}\kets{1}{B}\kets{0}{C}$ is a basic state of qubits $A$,
$B$ and $C$. We use the convention $\kets{010}{ABC} \defeq
\kets{0}{A}\kets{1}{B}\kets{0}{C}$.  Qubit labels are omitted when
they can be inferred from the context.  The (pure) state space of a
collection of qubits consists of the unit complex linear combinations
(called {\em superpositions\/}) of their basic states.

Quantum gates act on qubits by applying a unitary operator to the
current state.  An example is the NOT gate, which in matrix form is
given by the Pauli matrix $\sigma_x$. The NOT gate applied to qubit
$A$ is denoted by $\slb{\sigma_x}{A}$ and has the effect of flipping
the binary label associated with $A$ in the basic states. The effect
on superpositions is obtained by linear extension.

To describe gates and their effects we can use the bra-ket
conventions. In addition to the ket
symbols already introduced, we introduce {\em bra\/} symbols $\bras{b}{X}$ for
qubit $X$ with $b=0$ or $b=1$. Formal linear combinations of bra and ket
symbols can be multiplied using distributivity and associativity
rules together with the following:
\begin{itemize}
\item[1.] Bras and kets with different labels commute.
\item[2.] $\bras{a}{X}\kets{b}{X} = \delta_{a,b}$.
\item[3.] Expressions involving two kets or two bras
with the same label next to each other are illegal.
\end{itemize}
If $\phi$ is a bra-ket expression, then so is $\phi^\dagger$, which is
obtained by conjugating the complex coefficients, reversing the order
of elementary products and changing kets into bras and vice-versa.
For example $\kets{0}{A}^\dagger = \bras{0}{A}$.

With these conventions, we can write the NOT gate acting
on qubit $A$ as
\begin{eqnarray}
\sigma_x^{A} &=& \kets{0}{A}\bras{1}{A}+\kets{1}{A}\bras{0}{A},
\end{eqnarray}
where $\sigma_x^{A}$ is intended to be applied to a state by
multiplication on the left.  The elementary gates available to a QRAM
are unitary operators acting on one or two qubits. The operator $U$ is
unitary if $U^\dagger U$ acts as the identity. Note that in the bra-ket
notation, there are many ways of writing the identity
operator. Examples include
\begin{eqnarray}
1 &=& \kets{0}{A}\bras{0}{A} + \kets{1}{A}\bras{1}{A}\\
  &=& \sum_b\kets{b}{AB\ldots}\bras{b}{AB\ldots}.
\end{eqnarray}
The elementary gates to be used here are based on exponentials
of products of the Pauli operators $\sigma_x$ and
\begin{eqnarray}
\sigma_y &=& -i\ket{0}\bra{1}+i\ket{1}\bra{0}\\
\sigma_z &=& \ket{0}\bra{0}-\ket{1}\bra{1}.
\end{eqnarray}
For qubits labeled by $1,\ldots,n$, a general
product of Pauli operators is denoted by $\sigma_b$,
where $b$ consists of $n$ pairs of bits and
is defined by
\begin{eqnarray}
\sigma_b &=& \prod_{i=1}^{n}\slb{\sigma_{b_i}}{i},
\end{eqnarray}
with the conventions $\sigma_{00}\defeq I$, $\sigma_{01}\defeq
\sigma_x$, $\sigma_{11} \defeq \sigma_y$ and $\sigma_{10} = \sigma_z$.
The {\em weight\/} of $\sigma_b$ is the number of non-zero pairs of
bits in $b$.  A {\em rotation by $\theta$ around $\sigma_b$\/} is the
operator
\begin{eqnarray}
e^{-i\sigma_b\theta/2} &=& \cos(\theta/2) - i\sin(\theta/2)\sigma_b.
\end{eqnarray}
A complete set of one and two qubit gates can be obtained from the set
of rotations by $\pm 2\arccos(4/5)$ around operators of weight at most
two~\cite{bernstein:qc1997a,knill:qc1997a,knill:qc1998c}.  A
polynomially equivalent model is obtained by allowing such rotations
around any product of Pauli operators. We adopt this model.

A general QRAM may at any time {\em measure\/} a qubit and act
according to the measurement outcome.  Suppose the state of the
qubits is given by $\psi$. Suppose the QRAM measures qubit
$A$. In bra-ket notation we can expand $\psi=\kets{0}{A}\psi_0 +
\kets{1}{A}\psi_1$, with $\psi_0$ and $\psi_1$ not
containing any kets labeled $A$.  Let $p=\psi_0^\dagger\psi_0$ and
$q=\psi_1^\dagger\psi_1$.  Then $p$ and $q$ are positive reals with
$p+q=1$.  The effect of the measurement projects the qubits into the
state $\kets{0}{A}(1/p)\psi_0$ with probability $p$ and into the state
$\kets{1}{A}(1/q)\psi_1$ with probability $q$.  The {\em answer\/} of the
measurement is $0$ in the former case, and $1$ in the latter, and the
answer is placed into a (classical) bit register.
We simplify this model by permitting only measurements
of qubit $1$ and assuming that all the qubits used so far are lost after
the measurement. This simplified model is polynomially equivalent
to the general one with respect to
bounded error algorithms for promise problems.

A version of the one-bit model of quantum computation is given by the
Q1RAM, which differs from the (simplified) QRAM only in that the
initial state of the qubits has qubit $1$ in state $\ket{0}$ and all
the other qubits in a state picked uniformly at random from the basic
states.  A measurement of qubit $1$ also re-initializes the qubits.
Surprisingly, there are problems for which no efficient classical
algorithm is known and that can be solved efficiently using a Q1RAM, while
Q1RAMs are not as powerful as QRAMs with respect to
oracles~\cite{knill:qc1998c}.

\section{Simulating Quantum Computers}

As described above, both models of quantum computation can be thought
of as being based on classical deterministic RAMs with access to
certain oracles. The input to the oracles is a sequence of quantum
gates and the answer is $0$ or $1$ with the appropriate probability
distribution.  A fundamental question is whether a probabilistic RAM
can efficiently implement these oracles. Note that the output
probability distribution in such an implementation can deviate from
the correct one by $O({\epsilon/ N})$, where $N$ is the total number
of oracle calls, without significantly affecting the output of an
algorithm.

The problems solved by the oracles can be cast 
in terms of promise problems. In particular, the following
promise problems can be solved efficiently by quantum computers
and one bit quantum computers, respectively:

\begin{Problem}
Given a quantum network and the promise that after applying
the quantum network to the initial state $\ket{00\ldots}$,
the probability $p$ that the first qubit is in state $\ket{1}$
satisfies $|2p-1| \geq 1/2$, determine the sign
of $2p-1$.
\label{problem:net_est}
\end{Problem}

\begin{Problem}
Given a quantum network and the promise that after applying
the quantum network to the initial state with the first
qubit in state $\ket{0}$ and the others random,
the probability $p$ that the first qubit is in state $\ket{1}$
satisfies $|2p-1| \geq 1/2$, determine the sign
of $2p-1$.
\label{problem:net1_est}
\end{Problem}

\begin{Theorem}
A probabilistic RAM with access to an oracle for
Problem~\ref{problem:net_est} can efficiently simulate
a quantum computer.
\label{theorem:net_est=qc}
\end{Theorem}

We do not know whether a similar theorem
holds for the one-bit model of quantum computation
with respect to Problem~\ref{problem:net1_est}.

\proof
Suppose that we are given a quantum network $\cG$. 
The goal is to produce a random bit with probability
distribution close to the output qubit's distribution for $\cG$.
The first step is to use an oracle for Problem~\ref{problem:net_est}
to estimate the probability that the output qubit is in state $\ket{1}$.
To do so we design new quantum networks $\cG_{x,N}$.
$\cG_{x,N}$ applies $\cG$ to $(N/\epsilon)^2$ independent sets of qubits,
then uses ancillas to (reversibly) determine whether
the fraction of $\ket{1}$'s in the $(N/\epsilon)^2$ output qubits is greater
than $x$ or not, placing the answer into its output qubit.
The oracle is queried for $\cG_{x,N}$. By using binary search
on $x$, the desired probability can be determined to
within $O({\epsilon/ N})$ in $O(\log(N/\epsilon))$ queries.
The probabilistic RAM then simulates the output of the quantum network by
producing a random bit with this estimated bias.
\qed

\section{Reduction to Quadratically Signed Weight Enumerators}

For a quantum network $\cG$, let $U(\cG)$ be the unitary operator
defined by $\cG$. Observe that without loss of generality, we can
restrict $\cG$ to have only real
gates~\cite{bernstein:qc1997a}. (Other networks can be simulated by
real networks using one ancilla qubit to keep track of phases, see
Appendix~\ref{appendix:real}.) These are gates involving rotations
around $\sigma_b$'s with an odd number of factors of the form
$\sigma_y$. The gate set is still complete if we assume also that the
orientation of the rotation is positive if the number of $\sigma_y$ is
$1\;\mod(4)$ and negative otherwise.

The results and arguments in~\cite{knill:qc1998c} show that
Problems~\ref{problem:net_est} and~\ref{problem:net1_est} are
equivalent to problems of estimating specific coefficients of an
operator representation of $U(\cG)$.
In particular, for networks with real gates only,
they correspond to the following two problems:

\begin{Problem}
Promise: $|\bra{00\ldots}U(\cG)\ket{00\ldots}| \geq 1/2$.
Determine the sign of $\bra{00\ldots}U(\cG)\ket{00\ldots}$.
\label{problem:ub_est}
\end{Problem}

\begin{Problem}
\label{problem:uo_est}
Let $n$ be the number of qubits used by $\cG$.
Promise: $|{1\over 2^n}\trace\, U(\cG)| \geq 1/2$.
Determine the sign of $\trace\, U(\cG)$.
\end{Problem}

Let $\cG$ be determined by the sequence of gates $G_1,\ldots,G_N$,
so that $U(\cG) = G_NG_{N-1}\ldots G_1$. Each
gate is of the form
\begin{eqnarray}
G_k &=& {4\over 5} \pm i {3\over 5} \sigma_{b_k},
\end{eqnarray}
where $b$ contains an odd number of pairs of the form $11$
and the sign ($\pm$) depends on the number of $\sigma_y$ in $\sigma_{b_k}$.
Let $|b|_y$ be the number of $\sigma_y$ occurring in $\sigma_b$
and define $\tilde\sigma_b = (-i)^{|b|_y}\sigma_b$.
Then, because of the condition on the signs of the rotations,
\begin{eqnarray}
G_k &=& {4\over 5} + {3\over 5}\tilde\sigma_{b_k}.
\end{eqnarray}
To expand the product of the $G_k$, we need to
determine the multiplication rules for the $\tilde\sigma_b$.
The property that $b$ has an odd number of pairs
of the form $11$  is defined by $b^TBb=1$,
where $B$ is block diagonal with
two-by-two blocks given by
\begin{eqnarray}
B_1 &=& \left(\begin{array}{rr}
             0&1\\0&0
            \end{array}\right).
\end{eqnarray}
Direct verification shows that the multiplication rules
are given by
\begin{eqnarray}
\tilde\sigma_{b_1}\tilde\sigma_{b_2} = (-1)^{b_1^TBb_2}\tilde\sigma_{b_1+b_2},
\end{eqnarray}
where the sum in the subscript is bit-by-bit, modulo two.
$U(\cG)$ can now be expanded as follows:
\begin{eqnarray}
U(\cG) &=& \prod_{k=N}^1 G_k\\
 &=& \prod_{k=N}^1 (4 + 3\tilde\sigma_{b_k})/5\\
 &=& {1\over 5^N}\sum_{a} (-1)^{a^T\lowertr(H^TBH)a}4^{|a|}3^{N-|a|}\tilde\sigma_{Ha}.
\label{sum:net_gen}
\end{eqnarray}
The last step requires distributing the product over the sum and using
the multiplication rules for the $\tilde\sigma_b$ operators.  $H$ is
the matrix whose columns are the $b_k$. The sum is over all $0$-$1$
column vectors $a$ of dimension $N$. The bits of the vector $a$
correspond to which of the two terms of each sum in the product are
chosen to get a summand of the expansion. The first bit of $a$
determines the term of the factor $G_1$, and so on. Note that every
matrix $H$ of dimension $2n$ by $N$ with the property
that $\diag(H^TBH) = I$ can occur in this expression.  The
coefficients of $U(\cG)$ to be estimated in
Problems~\ref{problem:ub_est} and~\ref{problem:uo_est} are
\begin{eqnarray}
\bra{00\ldots}U(\cG)\ket{00\ldots} &=& 
{1\over 5^N}\sum_{a:BHa=0} (-1)^{a^T\lowertr(H^TBH)a}4^{|a|}3^{N-|a|}\\
{1\over 2^n}\trace U(\cH) &=&
{1\over 5^N}\sum_{a:Ha=0} (-1)^{a^T\lowertr(H^TBH)a}4^{|a|}3^{N-|a|}.
\end{eqnarray}
Here the condition $BHa=0$ means that $\sigma_{Ha}$ has no
$\sigma_x$ or $\sigma_y$ factors.

It remains to obtain the simpler forms
of Problems~\ref{problem:estimate_dqc} and~\ref{problem:estimate_dqc1}.
Let $H_0$ and $H_1$ be the two $n$ by $N$ matrices obtained
from the even and the odd rows of $H$, respectively (starting
the count at zero, so that the first row is considered even).
The above sums are then equivalent to
\begin{eqnarray}
\bra{00\ldots}U(\cG)\ket{00\ldots} &=& 
{1\over 5^N}\sum_{a:H_1a=0} (-1)^{a^T\lowertr(H_0^TH_1)a}4^{|a|}3^{N-|a|}
\label{sum:dqc}\\
{1\over 2^n}\trace U(\cG) &=&
{1\over 5^N}\sum_{a:H_1a=0, H_0a=0} (-1)^{a^T\lowertr(H_0^TH_1)a}4^{|a|}3^{N-|a|}.
\label{sum:dqc1}
\end{eqnarray}
Any pair of matrices $H_0$ and $H_1$ with the property
that $\diag(H_0^TH_1)=I$ is possible in these sums.
To show that the sums of Problems~\ref{problem:estimate_dqc} and~\ref{problem:estimate_dqc1} are of this form, consider
first the case $k=4$ and $l=3$. The two sums can then be written as
\begin{eqnarray}
\sum_{a:Ca=0} (-1)^{a^T\lowertr(C) a} 4^{|a|}3^{N-|a|}\\
\sum_{a:Ca=0,C^Ta=0} (-1)^{a^T\lowertr(C) a} 4^{|a|}3^{N-|a|}.
\end{eqnarray}
In the former case, let $H_0 = I$ and $H_1 = C$, to see that it is an
instance of Sum~(\ref{sum:dqc}). (The factor of $5^N$ is properly
taken care of by the conditions in the promise.)  In the latter case,
observe that one can write $C=XY^T$ with $X$ and $Y$ rectangular
matrices with independent columns. This can be done by first using
Gaussian elimination to write $UCV^T=I_k$, where $U$ and $V$ are
invertible and $I_k$ is a partial identity matrix with $k$ ones, then
using such a decomposition for $I_k$. To see that this sum is in fact
an instance of Sum~(\ref{sum:dqc1}), let $H_0=X^T$ and $H_1 = Y^T$ and
observe that $H_0^TH_1a = 0$ iff $H_1a=0$ and similarly for
$H_1^TH_0$.

For other $k$ and $l$, use the above reductions to get sums like those
of~(\ref{sum:dqc}) and~(\ref{sum:dqc1}), but with $k$ and $l$
substituted for the numbers $4$ and $3$, respectively, and
$\sqrt{k^2+l^2}$ substituted for the divisor $5$. These sums
correspond to sums involving gates with different rotation angles.  By
universality, these gates can be approximated to within
$O(\epsilon/N)$ using the standard ones with
$\mbox{polylog}(N/\epsilon)$ overhead in
gates~\cite{solovay:qc1998a,kitaev:qc1997a}. There is a classical
algorithm that computes such approximations efficiently. The resulting
gate network can be turned back into a sum of the desired form.

To see that Sum~(\ref{sum:dqc}) can be cast in the form required by
Problem~\ref{problem:estimate_dqc} requires more work. Let $H_0$ and
$H_1$ be as in Sum~(\ref{sum:dqc}). If $H_0$ has independent rows,
then the constraint $H_1a=0$ is equivalent to $H_0^TH_1a=0$, so the
sum is of the desired form.  If not,
it is necessary to modify $H_0$ so that it has full rank
without changing the value of the sum.

\begin{Lemma}
There exists a full rank $H_2$ such that 
$\lowertr(H_2^TH_1) = \lowertr(H_0^TH_1)$ and $\diag(H_2^TH_1)=I$.
\end{Lemma}

\proof Consider the first $n$ columns of $H_0$ and $H_1$, labeled
$c_1,\ldots,c_n$ and $d_1,\ldots,d_n$ respectively.  To obtain
$H_2$, the $c_i$ are replaced by independent $c'_i$.  In order for
the desired equality to hold, we need $d_i^Tc'_j = d_i^Tc_j$ for
$i\leq j$. The desired $c'_j$ can be constructed starting with $c'_n$.
Let $c'_n$ be any solution to $d_i^Tc'_n = d_i^Tc_n$ for all $i\leq
n$. Such a solution exists and is non-zero because $d_n^Tc_n = 1$.
Suppose $c'_n,c'_{n-1},\ldots,c'_{k+1}$ have been constructed.  The
set of solutions to $d_i^Tx = d_i^Tc_k$ for $i\leq k$ is an affine
subspace not containing $0$ of dimension at least $n-k$. Its
intersection with the complement of the span of the
$c'_n,c'_{n-1},\ldots,c'_{k+1}$ is therefore not empty. Let $c'_k$ be
an element of this intersection. Proceed until $c'_1$ has been
obtained. The vectors constructed by this method satisfy the desired
conditions. \qed

For the matrix shown to exist by this lemma,
\begin{eqnarray}
\sum_{a:H_1a=0} (-1)^{a^T\lowertr(H_2^TH_1)a}4^{|a|}3^{N-|a|}
 &=&
  \sum_{a:H_1a=0} (-1)^{a^T\lowertr(H_0^TH_1)a}4^{|a|}3^{N-|a|}.
\end{eqnarray}
Thus, the constraint in the sum can be replaced
by $H_2^TH_1a=0$ to obtain a sum of the desired form.
We have proved the following:

\begin{Theorem}
Problem~\ref{problem:estimate_dqc} is polynomially equivalent
to Problem~\ref{problem:net_est}.
\label{theorem:dqc-red}
\end{Theorem}

\begin{Corollary}
Probabilistic
RAMs with an oracle for Problem~\ref{problem:estimate_dqc}
are polynomially equivalent to quantum computers.
\label{corollary:dqc=estimate_dqc}
\end{Corollary}

\begin{Corollary}
Problem~\ref{problem:estimate_dqc} is complete for BQP.
\label{corollary:estimate_dqc=bqp}
\end{Corollary}

\begin{Theorem}
Problem~\ref{problem:estimate_dqc1} can be solved efficiently
by one-bit quantum computers.
\label{theorem:dqc1-red}
\end{Theorem}

It is an open problem to determine whether the converse of
Theorem~\ref{theorem:dqc1-red} holds and to determine
the relationships between the various promise problems
suggested in the Introduction. Note that
it is possible to simulate one-bit quantum computers
given access to oracles for Problem~\ref{problem:estimate_dqc1},
if the coefficient $1/2$ in the bound in the promise 
is replaced by $1/X$, with $X$ given as an input.
($X$ should be given as a unary number to maintain canonical
size/complexity relationships.)

\section{Conclusion}

We have shown that the problem of simulating a
quantum algorithm on a classical computer is equivalent
to the difficulty of estimating certain combinatorial sums given by
the quadratically signed weight enumerators.  
The problem of approximating these sum includes a new set
of apparently difficult problems solvable efficiently by quantum
computers. The class of known problems of this type is still sparse.
Except for the ones proposed here, they are generally
related to finding periodicities in functions or inferring properties
of eigenvalues of unitary operators. Shor's factoring and discrete
logarithm algorithms are of this type~\cite{shor:qc1995a}. The
factoring and discrete logarithm problems have the advantage of not
requiring a potentially difficult to verify promise. On the other
hand, promise problems are a natural framework to use for both
probabilistic and quantum computation and abstract the much more
economically significant statistical inference problems underlying
many practical applications.  Our work demonstrates that quadratically
signed weight enumerator problems are both simple to state and have
sufficient flexibility to represent the capabilities of both quantum
computers and one-bit quantum computers. There are variants that
appear to be hard, perhaps even for quantum computers, and others
that may be easier than one-bit quantum computation but hard for
classical computation.  As a result, the investigation of
this class of problems will contribute toward a better understanding
of classical and quantum complexity classes.

\phead{Acknowledgments.} We thank Sean Hallgren for helpful
discussions.  This work was supported by the Department of Energy,
under contract W-7405-ENG-36, and by the NSA.

%\bibliographystyle{plain}
%\bibliography{journalDefs,qc}

\appendix

\section{Real Gates are Equivalent to Complex Gates}
\label{appendix:real}

Let $\cG$ be a gate network consisting of the gates $G_N,\ldots,G_1$.
Introduce a new qubit, labeled $0$, to represent the complex
phase by the real orthogonal map
\begin{eqnarray}
R:(\alpha\kets{0}{\,0}+\beta\kets{1}{\,0})\ket{b} &\rightarrow&
  (\alpha+i\beta)\ket{b}.
\end{eqnarray}
Define $G'_k = \Re(G_k) -i\slb{\sigma_y}{0}\Im(G_k)$.
Then the $G'_k$ are real orthogonal and define a new
gate network $\cG'$. Note that each $G'_k$ can be
efficiently approximated using the elementary real gates.
The unitary operator defined by $\cG'$ satisfies
\begin{eqnarray}
U(\cG') &=& \Re(U(\cG))-i\slb{\sigma_y}{0}\Im(U(\cG)),
\end{eqnarray}
and $U(\cG) = RU(\cG')R^{-1}$. These relationships
can be used to simulate any network by a real network.

\end{document}